\documentclass[a4paper,UKenglish]{lipics-v2018}

\usepackage{siunitx}
\usepackage[noend]{algpseudocode}
\graphicspath{{./FIGURES/}}   
\usepackage{amsmath} 
\usepackage{amsfonts} 
\usepackage{amssymb} 
\usepackage{xspace}   
\usepackage{relsize} 
\usepackage{verbatim} 
\usepackage{enumerate} 
\usepackage{enumitem} 
\usepackage{centernot} 
\usepackage{adjustbox} 
\usepackage{multirow} 
\usepackage[normalem]{ulem} 
\usepackage{ifthen} 
\usepackage{stmaryrd}
\usepackage{mathtools}
\usepackage[fleqn]{cases}
\newcommand{\signsquare}[1]{|#1|(#1)}

\newcommand{\xstep}[1]{x^{({#1})}}
\newcommand{\xop}{x^{*}}
\newcommand{\maxinitialflow}{\psi_{\max}}
\newcommand{\partialderivative}[2]{\dfrac{\partial {#1}}{\partial {#2}}}

\newcommand{\mareksmargincomment}[1]%
    {{%
      \marginpar{{\tiny\begin{minipage}{0.5in}
                       \begin{flushleft}
                          {\color{red}MCh} {#1}
                       \end{flushleft}
                       \end{minipage}
                }}
    }}

\newcommand{\ignore}[1]{}

\newcommand{\braced}[1]{{ \left\{ #1 \right\} }}

\newcommand{\bracked}[1]{{ \left[ #1 \right] }}
\newcommand{\bigbracked}[1]{{ \Big[\, #1\, \Big] }}

\newcommand{\parendbig}[1]{{ \big( \, #1 \, \big) }}
\newcommand{\parendBig}[1]{{ \Big( \, #1 \, \Big) }}

\newcommand{\suchthat}{{\;:\;}}

\newcommand{\absvalue}[1]{{| \, #1 \,|}}

\newcommand{\absvalueBig}[1]{{ \Big| \, #1 \, \Big| }}
\newcommand{\norm}[1]{{ \lVert #1 \rVert}}

\newcommand{\bart}{\bar{t}}

\newcommand{\barep}{{\bar{\varepsilon}}}

\newcommand{\myparagraph}[1]{{\medskip\noindent\textbf{#1}.}}

\newcommand{\reals}{{\mathbb{R}}}
\newcommand{\integers}{{\mathbb Z}}

\newcommand{\half}{{\textstyle{\frac{1}{2}}}}

\newtheorem{proposition}{Proposition}

\newtheorem{claim}{Claim}

\nolinenumbers
\title{A Note on Local Convergence of Iterative Processes for Pipe Network Analysis}

\author{Huong Luu}{Department of Computer Science\\ University of California at Riverside}{}{}{}

\author{Marek Chrobak}{Department of Computer Science\\ University of California at Riverside}{}{}{}

\Copyright{The copyright is retained by the authors}

\authorrunning{H.~Luu and M.~Chrobak}

\subjclass{%
}


\funding{Research supported by NSF grant CCF-1536026.}

\begin{document}

\maketitle

\begin{abstract}
Analysis of pipe networks involves computing flow rates and pressure differences on pipe segments in the
network, given the external inflow/outflow values. This analysis can be conducted using 
iterative methods, among which the algorithms of Hardy Cross and Newton-Raphson have historically been applied in practice. 
In this note, we address the mathematical analysis of the local convergence of these algorithms.
The loop-based Newton-Raphson algorithm converges quadratically fast, and we provide estimates
for its convergence radius to correct some estimates in the previous literature.
In contrast, we show that the convergence of the Hardy Cross algorithm is only linear. 
This provides theoretical confirmation of experimental observations reported earlier in the literature. 
\end{abstract}


\section{Introduction}
\label{sec: introduction}


The analysis of pipe networks is a well-studied problem in civil engineering that arises in the design and modeling of
large-scale hydraulic networks, such as water municipal systems, sewage systems, or substances distribution networks.
In the most common formulation, the objective is to determine 
the flow rates and pressure differences in all pipe segments within the network, given the its topology, nodal inflows/outflows, and
the parameters of the pipes and fluid (diameter, friction, fluid density, etc.)~ \cite{stephenson_pipeflow_1984}. 
At a fundamental level, this problem is analogous to the analysis of electrical circuits. 
These two problems share two similar physical principles: mass (or flow) conservation and energy conservation.
The law of flow conservation states that at each node, the total inflows and outflows are equal;
and the principle of conservation of energy requires that around a closed loop, the total sum of pressure differences is zero.
However, in contrast to their electrical counterparts, in large-scale pipe networks, 
the equation governing the relationship between flow and pressure difference in a pipe segment is non-linear\footnote{%
This non-linearity is due to fluid turbulence. In small-scale networks, such as those that arise in microfluidics, flows tend to be laminar and their behavior can be modeled with linear equations.}. %
This equation, known as the Darcy-Weisbach equation, asserts that the 
pressure difference in a pipe segment $e$ is proportional to the square of its flow: $|\Delta_e p| = \mu_e q_e^2$, 
where $|\Delta_e p|$ is the positive pressure difference, $q_e$ is the flow value, and $\mu_e$ 
is a constant coefficient that is inversely proportional to the pipe's diameter, and proportional to the friction factor, pipe length, and fluid density.

Pipe network analysis boils down to solving the system of the above-mentioned equations,
namely the flow conservation equations for all nodes and the energy conservation equations for all pipe segments.
The specific choice of variables, be it pressure values or flow values, leads to different formulations of these
equations, with three common approaches known as the loop method, the flow method, and the node method~\cite{boulos_comprehensive_water_distribution_2006}.
(See Section~\ref{sec: preliminaries} and Appendix~\ref{sec: oscillation example}.)
In either formulation, the resulting system of equations is non-linear, and computing its exact solution is not feasible.
Instead, iterative techniques that compute approximate solutions have been widely adopted in the scientific literature and in practice.


\myparagraph{Past work}
One of the earliest approaches to pipe network analysis was proposed by Cross in 1936~\cite{cross_analysis_1936}.
He introduced both the loop and node variants of his method\footnote{%
H.~Cross referred to these approaches, respectively, as the methods of balancing flows and balancing heads.},
that we denote \emph{HC-loop} and \emph{HC-node}, respectively.
Due to its simplicity, Algorithm~HC-loop has been used to analyze small networks with spreadsheets or hand calculations.
Another frequently used approach involves applying the generic Newton-Raphson algorithm for solving non-linear equations.
This algorithm can be applied to any of the above three settings: flow, loop or node, and we use notations
\emph{NR-flow}, \emph{NR-loop} and \emph{NR-node} for its respective variants.

In this work we focus mainly on algorithms based on the loop method. 
This method assumes that some initial flow satisfying the flow conservation conditions is given,
and its unknowns are the flow adjustments along the network's cycles with respect to this initial flow. 
(See Section~\ref{sec: preliminaries} for a detailed description).

While there is a fair amount of literature on experimental performance of these algorithms, relatively little is known about their provable convergence properties.
In~\cite{cross_analysis_1936}, Cross claimed that Algorithm~HC-loop's convergence is ``sufficiently rapid'', though this claim was not supported by rigorous analysis.
Adams~\cite{adams_flow_analysis_1955} provided an analysis of the accuracy of the flow corrections generated by Algorithm~HC-loop by comparing it with the exact correction values. 
This result is useful for convergence analysis of simple systems that consist of only one cycle; however, for larger systems, the mathematical analysis is incomplete. 
Empirical results in~\cite{lopes_implementation_Hardy_Cross_2004} show fast convergence of HC-loop for several networks in the dataset. 
Altman and Boulos report in~\cite{altman_convergence_Newton_nonlinear_1994} an attempt to analyze local convergence of NR-flow;
 however, as we explain in Appendix~\ref{sec: altman errors}, their analysis is invalid.
In~\cite{brkic_Improvement_of_Hardy_Cross_Method_2009}, Brkic conducted a comparative study of NR-loop and HC-loop, and his experimental results 
support the earlier observations that NR-loop exhibits faster convergence. 


\myparagraph{Our results}
We present an analysis of the local convergence of Algorithms~NR-loop and HC-loop in Section~\ref{sec: analysis of Newton method} and Section~\ref{sec: analysis of Hardy Cross method}.
We show that Algorithm~NR-loop converges quadratically fast if the initial flow is sufficiently close to the solution, and we provide a bound on the radius of its quadratic convergence.
The quadratic convergence of the Newton-Raphson method is not surprising --- in fact it is known to converge quadratically under some fairly mild assumptions.
Our aim here was to correct and refine the analysis of the convergence radius provided 
in paper~\cite{altman_convergence_Newton_nonlinear_1994}.

For Algorithm~HC-loop, we prove a similar local convergence property, although in this case the convergence rate is only linear. 
We also show that this bound is tight by exhibiting an example where Algorithm~HC-loop's convergence is not better than linear.
These results provide a theoretical confirmation of experimental observations, discussed earlier, about the superiority of NR-loop over HC-loop.
 

\myparagraph{Other related work}
As for other methods, Shamir and Howard commented in~\cite{shamir_water_distribution_analysis_1968,shamir_engineering_analysis_1977} that 
Algorithm~NR-node might not converge if it encounters a singular derivative matrix or if it enters and infinite loop, although no specific examples were given there. 
In Appendix~\ref{sec: oscillation example} we provide an example on which Algorithm~NR-node enters an infinite looop. 
A fairly extensive empirical comparison of different formulations of both Algorithm~HC and Algorithm~NR was presented by Wood~\cite{wood_algorithms_pipe_analysis_1981}. 
The results indicate that the loop methods encounter fewer convergence issues than the respective node methods, and that the Newton-Raphson algorithm is more reliable for all formulation variants.


\section{Preliminaries}
\label{sec: preliminaries}


\myparagraph{Flow graphs}
A pipe system can be represented by an undirected graph $G = (V,E)$, where $|V| = n$, $|E| = m$. 
In $G$, the pipe junctions are represented by vertices and pipe segments by edges. 
We assume that $G$ is biconnected, as otherwise its biconnected components can be analyzed separately. 
For notational reasons, it will be convenient to assign orientations to edges. 
To this end, we assume that $V = \braced{1,2,...,n}$, 
and we define the orientation of an edge $e$ between two different vertices $u$ and $v$
as being from $u$ to $v$ if $u < v$, and from $v$ to $u$ otherwise. 
For simplicity, the edges in $E$ will be identified by numbers $1,2,...,m$.
Let  $D$ be the $m \times n$ \emph{incident matrix} where $D_{ev} = 1$ 
if edge $e$ is incident with vertex $v$ and is oriented toward $v$, $D_{ev} = -1$ if $e$ is oriented away from $v$, 
and $D_{ev} = 0$ if $e$ is not incident with $v$. 
The specification of a flow graph also involves the \emph{consumption vector} $w \in \reals^{n}$ that satisfies  $\sum_{v=1}^{n} w_{v} = 0$. 
This vector represents external inflows and outflows, that is the boundary conditions of the flows in the network.


\myparagraph{Cycles}
A closed loop in the pipe network corresponds to a cycle in its flow graph $G$.
More generally, we define a \emph{cycle} in $G$ as an even-degree subgraph of $G$. 
A cycle is called \emph{simple} if it is connected and all of its vertices have degree two. 
If we view cycles as vectors in $\integers_2^{m}$ ($\integers_2$ is the finite field of two elements 0 and 1), 
the set of all cycles in $G$ forms a linear space over $\integers_2$ with vector addition being the operation of symmetric difference.  
The dimension of this space is $k = m-n+1$. 
A \emph{cycle basis} in $G$ is a collection of $k$ linearly independent simple cycles.
Each cycle in $G$ can be obtained as a linear combination of the cycles in the basis. 
For our analysis, we assume that an arbitrary cycle basis $C$ is provided for the flow graph $G$.
The \emph{total length of $C$}, denoted $\ell_C$, is defined as the sum of the lengths of its cycles.

Each simple cycle in a basis $C$ has two possible orientations, which we refer to as \emph{clockwise} and \emph{counter-clockwise}. 
These orientations are determined by the two cyclic orderings of vertices in the cycle. The clockwise orientation can be chosen arbitrarily and independently for 
each cycle in the basis.
An edge $e$ on cycle $c$ is called \emph{clockwise on $c$} if its orientation agrees with the clockwise orientation of $c$, and otherwise $e$ is called \emph{counter-clockwise on $c$.}
It is worth noting that an edge can be part of multiple cycles and may be clockwise in one cycle while being counter-clockwise in another.

Given a cycle base $C$, we will number its cycles $1,2,...,k$.
We can then define the \emph{edge-cycle matrix} $A$ of size $m \times k$ representing $C$. 
The value $A_{ec} \in \braced{-1,0,1}$ indicates the relationship between the orientations of edge $e$ and cycle $c$. 
Specifically, it is $1$ if edge $e$ is clockwise on cycle $c$, $-1$ if $e$ is counter-clockwise, and $0$ if $e$ is not part of $c$. 


\myparagraph{The flow method} 
A flow in $G$ is represented by a vector $q \in \reals^{m}$, with $q_e$ denoting the flow on edge $e$.
The sign of $q_e$ indicates the flow direction: $q_e > 0$ iff the flow direction of $q$ agrees with the orientation of $e$ and vice versa.
The goal of the flow method is to compute a flow $q$ that satisfies:
\begin{numcases}{}
	D^{\top} \, q + w \;=\; 0 \label{eq: flow conservation}
	\\
	A^{\top} \, U \, q	\;=\;0 \label{eq: energy conservation}
\end{numcases}
where $U = \text{diag}(|q_e|)$ for $e = 1,2, \cdots, m$, that is, $U$ is the diagonal matrix whose entries are the absolute values of the flows.

Equation~(\ref{eq: flow conservation}) is the flows conservation, and
Equation~(\ref{eq: energy conservation}) is the energy conservation. 
For simplicity, we assume that the coefficient $\mu_e$ in the Darcy-Weisbach equations is equal $1$ for all edges. 
However, our derivation can be easily adapted to arbitrary coefficients.
With this setup, the pressure difference $\Delta_e p$ along an edge $e$ can be written as $\Delta_e p = |q_{e}| \, q_e$. 
Equation~(\ref{eq: energy conservation}) states that the pressure differences along each cycle need to add up to $0$. 
The energy conservation principle applies to all cycles in $G$, although for our purposes, it is sufficient to include only the $k$ equations for cycles in the given cycle basis $C$. 


\myparagraph{The loop method} 
The loop method is fundamentally equivalent to the flow method. 
In the loop method, the flow conservation equations~(\ref{eq: flow conservation}) are eliminated by the introduction of an arbitrary 
\emph{reference flow} $\psi \in \reals^{m}$ that already satisfies this equation.
Then, the flows in $G$ can be expressed as: 
\begin{equation}
	\label{eq: reference flow}
	q(x) \;=\; \psi + A \,  x,
\end{equation}
where $x \in \reals^{k}$ is a variable whose value represents the flow adjustments along cycles in $C$.
Thus the flow on an edge $e$ is $q_e(x) = \psi_e + \sum_{c=1}^k A_{ec}x_c$.
In other words, along each cycle $c$,  $x_c$ is added to flows on clockwise edges and subtracted from flows on counter-clockwise edges.
In this setting, we define the error function $F: \reals^k \mapsto \reals^k $ to represent
the deviation from the energy conservation equations, and using this function we can express the loop method as: 
\begin{equation}
	\label{eq: system of error functions}
	F(x) \;=\; A^{\top} \, U(x) \, q(x)	\; =\; 0
\end{equation}
where $U(x)$ is the $m \times m$ diagonal matrix defined above for the flow $q(x)$.  That is, 
if $F(x) = [f_1(x),f_2(x),...,f_k(x)]^{\top}$ then
$f_c(x) = \sum_{e=1}^m A_{ec} \, q_e(x) |q_e(x)| = 0$, for each cycle $c = 1,2,...,k$ in the cycle basis $C$.

We assume that this system has a solution and that the solution is unique. Both existence and uniqueness can be
established by using the equivalence to the flow method, and reducing the
flow method equations to solving a convex optimization problem~\cite{singh_on_the_flow_problem_uniqueness_2020}.
In~\cite{singh_on_the_flow_problem_uniqueness_2020}, a different proof of uniqueness is also given.


\myparagraph{Iterative algorithms} 
Our focus is on the analysis of solving the non-linear system~(\ref{eq: system of error functions}) using two iterative algorithms
discussed in the introduction: NR-loop and  HC-loop. 
In general, an iterative algorithm starts with some initial candidate solution $\xstep{0}$, and then it
produces successive approximate solutions $\xstep{t}$, attempting to reduce the value of $\norm{F(\xstep{t})}$.
The process repeats until some stopping criteria are satisfied. Such criteria may involve, for example, an upper bound on the
maximum number of iterations, or the improvement being below a desired value.
This improvement can be measured either by $\norm{\xstep{t} - \xstep{t-1}}$ or $\norm{ F(\xstep{t}) - F(\xstep{t-1})}$.

An iterative algorithm is called \emph{convergent} if it generates a sequence of values that converges to the actual
solution, that is $F(\xstep{t}) \xrightarrow[t]{} 0$, if it's allowed to run indefinitely.
The efficiency of an iterative process can be measured by its rate of convergence, which captures how quickly the generated sequence approaches the solution. 
The sequence $\braced{\xstep{t}}$ is said to \emph{converge to a solution $\xop$ with $\omega$-order}~\cite{jay_a_note_on_q_convergence_2001}, 
for some $\omega \ge 1$,
if there exist two constants $\barep \ge 0$ and $\bart \ge 1$ such that for all $t \ge \bart$: 
\begin{equation*}
	\norm{\xstep{t} - \xop} \le \barep \, \norm{ \xstep{t-1} -  \xop }^{\omega}.
\end{equation*} 
For $\omega =1$ the convergence is said to be linear, and for $\omega=2$ it is called quadratic.


\section{Analysis of Newton-Raphson Algorithm}
\label{sec: analysis of Newton method}


The Newton-Raphson algorithm is an iterative process widely used for finding zeros of non-linear functions. 
For single-argument functions $\reals\to\reals$, the algorithm obtains a new approximation value as the intersection of the tangent line to its graph at the current value with the x-axis. 
This concept can be naturally extended to multi-variate functions $\reals^k\to\reals^k$ by using the $k\times k$ Jacobian matrix, which represents the first-order partial derivatives.

Given a flow graph $G$, a cycle basis $C$ of $G$ and a reference flow $\psi$, we apply the Newton-Raphson algorithm (NR-loop)
to solve the system of equations~(\ref{eq: system of error functions}) starting from some initial solution $\xstep{0}$. 
(Note that we can as well assume that $\xstep{0} = 0$, for otherwise we can add $\xstep{0}$ to the reference flow and 
set up the equations for this modified reference flow.)
At each step $t\ge 1$, the new solution $\xstep{t}$ is obtained from the previous one using the formula: 
\begin{equation}
	\label{eq: NR step function}
	\xstep{t} \;=\; \xstep{t-1} - F'(\xstep{t-1})^{-1} \, F(\xstep{t-1})
\end{equation}
where $F'(x)$ denotes the Jacobian of $F$ at $x$ which can be expressed compactly in matrix form as: 
\begin{equation}
	\label{eq: NR derivative function}
	F'(x) \;=\; 2 \, A^{\top} \, U(x) \, A
\end{equation}

It is interesting to note that from the setup, it might seem that the flows $q(\xstep{t})$ obtained by applying  the adjustment $\xstep{t}$ would be different for different cycle bases of $G$. 
However, upon closer examination, we find that it is not the case. 
This observation aligns with Nielsen's finding in~\cite{nielsen_methods_for_analyzing_pipe_networks_1989}. 
We offer a straightforward proof below. 
\begin{proposition}
\label{proposition: cycle independence}
	Given a flow graph $G$, a reference flow $\psi$ and and initial approximation $\xstep{0}$,
	the flows obtained at each step of Algorithm~NR-loop are independent of the choice of cycle basis for $G$. 
\end{proposition}
\begin{proof}
	Let  $A, B$ be two edge-cycle matrices that correspond to two different cycle bases of $G$.
	Using $A$ and $B$, set up the two error functions $F_A, F_B$ and let their respective Jacobians be $F'_A, F'_B$. 
	Denote by $\xstep{t}_A, \xstep{t}_B$ the adjustments at step $t$ acquired from applying Algorithm~NR-loop on $F_A, F_B$ respectively. 
	Setting $\xstep{0}_A = \xstep{0}_B = \vec{0}$, we first show that  $A \, x_A^{(1)} = B \, x_B^{(1)}$. 

	We can express $B = AW$ for some $k \times k$ change of basis matrix $W$. 
	Assuming that $F'_A(0)$ and $F'_B(0)$ are invertible, we have: 

	\begin{align*}
	 	B \,\xstep{1}_B \;&=\;  - B \, F'_B(0)^{-1} \, F_B(0)
		\\
		&=\;- \half\cdot B \, [B^{\top}\, U_B(0) \, B]^{-1} \, B^{\top}\,  U_B(0) \, \psi
		\\
		&=\;  - \half\cdot  A \, W \, [ W^{\top} \, A^{\top } \, U_A(0) AW]^{-1} \, W^{\top}  \, A^{\top} \, U_A(0) \, \psi
		\\
		&=\;  - \half\cdot  A \, W \, W^{-1} [A^{\top } \, U_A(0) A]^{-1} \, (W^{\top})^{-1} \, W^{\top}  \, A^{\top} \, U_A(0) \, \psi
		\\
		&=\; - \half\cdot  A \, [A^{\top } \, U_A(0) A]^{-1}  \, A^{\top} \, U_A(0) \, \psi
		\\
		&=\;  - A \, F'_A(0)^{-1} \, F_A(0) = A \, \xstep{1}_A 
\end{align*}

Since the new flows are the same, we have $U_A(\xstep{1}_A) = U_B(\xstep{1}_B)$ which implies that the new Jacobian $F'_B(\xstep{1}_B)$ is invertible if and only if $F'_A(\xstep{1}_A)$ is invertible. 
This completes the proof. 

\end{proof}

The convergence of the Newton-Raphson algorithm for non-linear systems has been studied extensively for decades. 
The rate of convergence can be as fast as quadratic if the function satisfies certain conditions, for
example, when the initial approximation is in close neighborhood of the root and
the Jacobians at all successive approximations are invertible.
However, the root itself being unknown, it is generally difficult to determine if the initial point is ``sufficiently'' close to it. 
The Kantorovich~\cite{Kantorovich_functional_analysis_1969} theorem circumvents this challenge by giving the 
local convergence conditions in terms of the initial point and some general properties of the function. 
Our statement of this theorem, given below, follows the formulation in~\cite{ortega_newton_kantorovich_theorem_1968} 
but is adapted to functions on Euclidean spaces with one unique root. 

For $s \in\reals^k$ and $r\in \reals$, by $B(s,r)$ we denote the ball centered at $s$ with radius $r$, that is
$B(s,r) = \braced{x \in\reals^k\suchthat \norm{x-s}\le r}$.

\begin{theorem}[Kantorovich]~\label{thm: kantorovich}
		Let $F:  \reals^k \mapsto \reals^k$ be a differentiable function, 
		$F'(x)$ be the $k \times k$ Jacobian matrix of $F(x)$, and $\xstep{0} \in \reals^k $ be an initial approximation of
		the Newton-Raphson process for $F(x) = 0$. 
		Assume that: 
		\begin{description}
			\item{\emph{(1)}} $F'(\xstep{0})$ is invertible and $\norm{F'(\xstep{0})^{-1}} \le \beta$
			\item{\emph{(2)}}  $\norm{F'(\xstep{0})^{-1} \, F(\xstep{0}) } \le \eta$
			\item{\emph{(3)}} $\norm{F'(x)-F'(y)} \le L \norm{x-y} \;\; \textrm{for all}\; x,y \in \reals^k$
		\end{description}
		for some $\beta, \eta, L > 0$.
			
		With these assumptions,
		if $ h = \beta \eta L < \half$ then the sequence $\braced{\xstep{t}}$ generated by 
		the Newton-Raphson's iteration process starting at $\xstep{0}$ is well-defined, contained in the ball $B( \xstep{0} ,r)$ for $r = {(1-\sqrt{1-2h})}/{(\beta L)}$, 
		and converges quadratically to the unique solution $\xop$  of $F(x)=0$.
\end{theorem}

Any induced matrix norm $\norm{\cdot}$ can be used in this theorem. 
In our work, we assume the infinity norm on $\reals^k$, that is $\norm{\cdot} = \norm{\cdot}_{\infty}$. 
Specifically, the norm of a vector $x \in \reals^k$ is $\norm{x} = \max_{1\le c \le k} \absvalue{x_c}$, and the
norm of a matrix $M \in \reals^{k \times k}$ is $\norm{M} = \max_{1 \le c \le k} \sum_{d=1}^{k} \absvalue{M_{cd}}$. 

Given a flow graph $G$, we can use Theorem~\ref{thm: kantorovich} to estimate the radius $r$ of quadratic convergence expressed in terms of properties of $F$ and attributes of $G$. 
To this end, we need to estimate the constants in (1), (2) and (3) from Theorem~\ref{thm: kantorovich}.
Computing $F(\xstep{0})$ is straightforward, while  the inverse of $F'(\xstep{0})$ requires more computation. 
However, since no provably accurate and efficient estimates for $\norm{F'(x)^{-1}}$ are known, we can use $\beta = \norm{F'(\xstep{0})^{-1}}$ and $\eta =  \norm{F'(\xstep{0})^{-1} \, F(\xstep{0})}$. 

Our attention is directed towards the bound~(3), which involves the Lipschitz condition on the Jacobian matrix $F'$. 
Even though the Newton steps are independent of the choice of cycle basis $C$,
as proven in Proposition~\ref{proposition: cycle independence}, the Lipschitz constant may take on different values 
depending on $C$.  Let $\ell = \ell_C$ be the total length of $C$. We present a general estimate in terms of $\ell$:

\smallskip
\begin{claim}
	\label{claim: general bound on L}
	$\norm{F'(x)-F'(y)} \le 2 k \, (\ell - k + 1)  \, {\norm{x-y}}, \;\;\text{for all}\; x,y \in \reals^k$.
\end{claim}

\begin{proof}
The derivation for this bound uses the formulas~\eqref{eq: NR derivative function} for the Jacobian of $F(x)$
and~\eqref{eq: reference flow} for the flow values. First, compute:
\begin{equation*}
 F'(x)_{cd}
		 \;=\;  \frac{\partial f_c}{\partial x_d} (x)
		 \;=\; 2 \sum_{e=1}^{m} A_{ec} \, A_{ed} \, \absvalue{ \psi_e + \sum_{i=1}^{k} A_{ei} \, x_i} 
\end{equation*}
Then, 
\begin{align*}
    	\norm{F'(x)-F'(y)}	
				\;&=\;   \max_{1 \le c \le k}  \sum_{d=1}^{k} \absvalueBig{\frac{\partial f_c}{\partial x_d} (x) -  \frac{\partial f_c}{\partial x_d} (y)} 
				\\
				 \;&=\;   \max_{1 \le c \le k}  \sum_{d=1}^{k} 2 \, \absvalueBig{\sum_{e=1}^{m} A_{ec} \, A_{ed} \parendBig{ \absvalue{ \psi_e + \sum_{i=1}^{k} A_{ei} \, x_i} 
											-   \absvalue{ \psi_e + \sum_{i=1}^{k} A_{ei} \, y_i} }} 
				\\
				 \;&\le\;   \max_{1 \le c \le k} \sum_{d=1}^{k} 2 \,  \sum_{e=1}^{m} \absvalue{A_{ec}} \, \absvalue{A_{ed}} \absvalueBig{\sum_{i=1}^{k}  A_{ei} \, (x_i - y_i)}
				 \\
				 \;&\le\; 2 \,  \max_{1 \le c \le k} \sum_{d=1}^{k} \sum_{e=1}^{m} \absvalue{A_{ec}} \,  \absvalue{A_{ed}} \, \sum_{i=1}^{k} \absvalue{ A_{ei}}  \, \norm{x-y}
				 \\
				 \;&\le\; 2 \, (\ell - k + 1) \, k \, \norm{x-y} 
\end{align*}
because $\sum_{i=1}^{k}  \absvalue{A_{ei}}  \le k $ and $\max_{1 \le c \le k} \sum_{d =1}^{k} \sum_{e=1}^{m} \absvalue{A_{ec}} \, \absvalue{A_{ed}} \le \ell - k + 1$,
with the last inequality following from the observation that
for each cycle $d \ne c$,  at least one of its edge is not on $c$. 
\end{proof}


We are now ready to state sufficient conditions for quadratic convergence of Algorithm NR-loop on the loop equation~\eqref{eq: system of error functions}. 
This condition applies to a flow graph $G = (V,E)$, a cycle basis $C$ of $G$ with total length $\ell$, a reference flow $\psi$, and an initial solution $\xstep{0}$ 
where $F'(x^{(0)})$ is invertible: 

\begin{theorem}[Convergence conditions of Algorithm~NR-loop]\label{thm: nr convergence}
Let  $\beta =  \norm{F'(\xstep{0})^{-1}}$, $\eta =  \norm{F'(\xstep{0})^{-1} \, F(\xstep{0})}$, and $L = 2k (\ell - k + 1)$. 
If $\beta \eta L < \half$, then Algorithm~NR-loop converges to the unique solution $\xop$ of $F(x) = 0$ with quadratic order. 
\end{theorem}

%

In what follows, we discuss more specific estimates for $L$. 
The estimate of the convergence radius in Theorem~\ref{thm: nr convergence} critically depends on the choice of the cycle basis.
Trivially, for any basis, we have $\ell \le kn$. 
Therefore, in Theorem~\ref{thm: nr convergence}, $\ell$ can be replaced by $kn$ if this bound is sufficient.
However, in general, using a cycle basis $C$ with small total length $\ell$ is likely to improve the convergence radius.
The problem of computing cycle bases with small total length has been well studied.
In~\cite{rizzi_minimum_weakly_fundamental_cycle_bases_2009}, Rizzi introduced an $O(mn)$-time algorithm that generates 
a cycle basis of total length $\ell = O(m \log n)$, which is within a $2\log n$ factor of the optimum length. 
This length bound was further refined to $\ell = O(m \log n/ \log (m/n))$ by Kaufmann and Michail, as mentioned in~\cite[Theorem 4.5]{kavitha_cycle_basic_survet_2009}. 
If the loop equations are built upon the aforementioned cycle basis, the enhanced Lipschitz constant $L$ becomes $2k (\tau m \log n/ \log (m/n) - k + 1)$ for some constant $\tau$.

For certain types of graphs, better estimates are possible. 
In particular, planar graphs emerge naturally when studying flows in pipe networks.
If $G$ is planar, the bound on the Lipschitz constant $L$ can be improved
by using the face cycle bases in which every edge belongs to at most two cycles (faces):
\begin{equation}
	L \;\le\;  8n.
	\label{eqn: L for planar graphs}
\end{equation}
Unlike the formula for $L$ in Theorem~\ref{thm: nr convergence}, the proof of~\eqref{eqn: L for planar graphs} does not follow from
Claim~\ref{claim: general bound on L} directly. Instead, we argue as follows:
For every edge $e$, $\sum_{i=1}^{k} \absvalue{A_{ei}} \le 2$. 
Furthermore, for every cycle $c$, $\sum_{d \ne c} \absvalue{\frac{\partial f_c}{\partial x_d}} \le \frac{\partial f_c}{\partial x_c}$. 
Thus, 

\begin{align*}
    	\norm{F'(x)-F'(y)}  \;&\le\;    \max_{1 \le c \le k}  2 \absvalueBig{\frac{\partial f_c}{\partial x_c} (x) -  \frac{\partial f_c}{\partial x_c} (y)}
	\\
	 \;&\le\; \max_{1 \le c \le k} 4 \,  \sum_{e=1}^{m} A_{ec}^2 \, \absvalueBig{\sum_{i=1}^{k}  A_{ei} \, (x_i - y_i)}
				\\
				 \;&\le\;  \max_{1 \le c \le k} 4	\, \sum_{e=1}^{m}A_{ec}^2 \, \sum_{i=1}^{k}  \absvalue{A_{ei}}  \, \norm{x-y}
				 \\
				\;&\le \;  8n \, \norm{x - y}  
\end{align*}
because $\sum_{e=1}^{m} A_{ec}^2 \le n$ for any simple cycle $c$. 

\myparagraph{Example}
We illustrate the local convergence conditions on the simple flow graph depicted in the figure below.
The graph is planar and has $4$ vertices, $6$ edges, $2$ inlets and $1$ outlet. 
The cycle basis that is used to set up the loop equations is the cycle basis consisting of internal faces.
The clockwise directions of the cycles are shown in this figure.  

\medskip
\begin{minipage}{0.45\textwidth}
	\includegraphics[width= \linewidth]{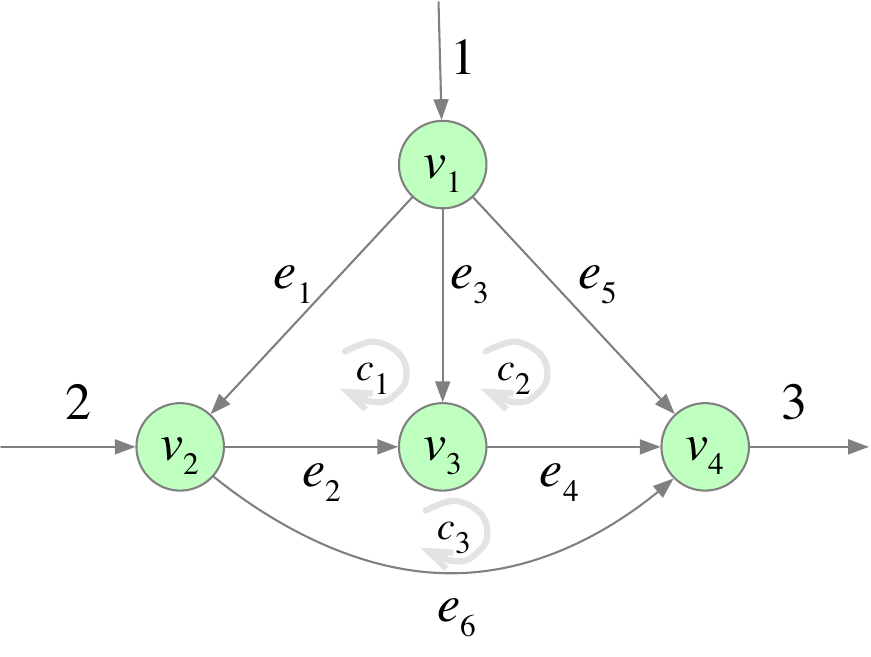}
\end{minipage}
\hfill
\begin{minipage}{0.45\textwidth}
\begin{equation*}
	\begin{bmatrix}
      		-1 & 0 &  0 \\
		-1 & 0 & 1 \\
		1 & -1 & 0 \\
		0 & -1 & 1 \\
		0 & 1 & 0 \\
		0 & 0 & -1
	\end{bmatrix}
\end{equation*}
\end{minipage}

Let the reference flow be  $\psi =  [1,1,0,1,0,2]$. The loop equations are: 
\begin{equation*}
	\begin{cases}
		f_1 (x) \;&=\; -\signsquare{1-x_1} - \signsquare{1-x_1+x_3} + \signsquare{x_1-x_2}
		\\
		f_2 (x) \;&=\;  -\signsquare{x_1-x_2}  - \signsquare{1-x_2+x_3}  + \signsquare{x_2}
		\\
		f_3 (x) \;&=\; \signsquare{1-x_1+x_3}  + \signsquare{1-x_2+x_3} - \signsquare{2-x_3}
	\end{cases}
\end{equation*}
and, the Jacobian is:
$F'(x) = \textstyle [ \frac{\partial F}{\partial x_1} (x) , \frac{\partial F}{\partial x_2} (x), \frac{\partial F}{\partial x_3} (x) ]$, where
%
\begin{align*}
	&\dfrac{\partial F}{\partial x_1} (x) = 2  \begin{bmatrix}
		|1-x_1| + |1-x_1+x_3| + |x_1-x_2|
		\\
		- |x_1-x_2|
		\\
		-|1-x_1+x_3|
	\end{bmatrix}
	\\
	&\dfrac{\partial F}{\partial x_2} (x) = 2  \begin{bmatrix}
	- |x_1-x_2|
	\\
	|x_1-x_2| +|1-x_2+x_3| + |x_2| 
	\\
	- |1 - x_2 + x_3|
	\end{bmatrix}
	\\
	&\dfrac{\partial F}{\partial x_3} (x) = 2  \begin{bmatrix}
	-|1-x_1+x_3|
	\\
	 - |1-x_2+x_3|
	\\
	 |1-x_1+x_3| + |1-x_2+x_3| + |2 - x_3|
	\end{bmatrix}
\end{align*}

Let the initial guess be $\xstep{0} = [1.38,1,0.93]$. It can be verified that $F'(\xstep{0})$ is invertible.
The constant in~(1) and (2) are $\norm{F'(\xstep{0})^{-1}} < 0.7 =  \beta$, $ \norm{F'(\xstep{0} )^{-1}\, F(\xstep{0})} < 0.005 = \eta$
As for the constant in (3), we can set $L = 8n = 32$ since the face cycle basis is used. 
This gives $ \beta \eta L < 0.12$, so the convergence condition is satisfied, and the system converges with quadratic order starting at the given $\xstep{0}$.


\section{Analysis of Hardy Cross Algorithm}
\label{sec: analysis of Hardy Cross method}

In order to simplify the calculation of the new solution $\xstep{t}$ in equation~\eqref{eq: NR step function}, 
a linear operator $H(x)$, typically associated with the Jacobian $F'(x)$, can be used in lieu of it: 
\begin{equation}
	\label{eq: linear op step function}
	\xstep{t} \;=\; \xstep{t-1}- H(\xstep{t-1})^{-1} \, F(\xstep{t-1})
\end{equation}

This variation of the Newton-Raphson method has been thoroughly explored in the literature. In the context of  pipe network analysis, 
Algorithm HC-loop can be viewed as a simplified version of the Newton-Raphson method where the Jacobian matrix $F'$ is replaced by the linear operator $H$, defined as follows:
\begin{equation*}
	H(x) \;=\; \text{diag} \bigbracked{\, \partialderivative{f_c}{x_c}(x) \,} \; \text{for}\, c = 1,2,\cdots,k
\end{equation*}
where $\partial f_c / \partial x_c$ are the diagonal entries of the matrix $F'(x)$, defined in equation~\eqref{eq: NR derivative function}. 

As the computation of $H(x)$ is relatively straightforward, Algorithm HC-loop was designed to support manual analysis of small-scale water distribution systems.  
The process also begins with a reference flow $\psi$ that satisfies the flow conservation equation~\eqref{eq: flow conservation}. 
It then iteratively produces new flow adjustments using equation~\eqref{eq: linear op step function} until the energy conservation equation~\eqref{eq: energy conservation} is satisfied, within some specified tolerance, for all cycles in a given cycle basis $C$ of $G$. 

It is important to note that although equation~\eqref{eq: linear op step function} may give the impression that the adjustments for all cycles need to be computed simultaneously, they can be performed for each cycle $c \in C$ separately as follows: 
\begin{align*}
	\xstep{t}_c  \;&=\; \xstep{t-1}_c - \bigbracked{\partialderivative{f_c}{x_c}(\xstep{t-1}) }^{-1} \, f_c(\xstep{t-1})  
				\;=\;  \xstep{t-1}_c  - \dfrac{\sum\limits_{e = 1}^{m} A_{ec} \, \signsquare{q_e^{(t-1)}}}
				{2\sum \limits_{e = 1}^{m} A_{ec}^2  \absvalue{q_e^{(t-1)}}} 
\end{align*}
where $q_e^{(t-1)} = \psi_e  + \sum_{i=1}^{k} A_{ei} \, \xstep{t-1}_i$ is the flow value on edge $e$ at step $t-1$. 

The difference between Algorithm HC-loop and Algorithm NR-loop lies in the fact that in Algorithm NR-loop
each adjustment takes into account the interaction between overlapping cycles, whereas in Algorithm HC-loop
these adjustments are computed independently for each cycle. 

\smallskip
Local convergence analysis of the variant of the Newton-Raphson algorithm, where a linear operator is used instead of the Jacobian matrix, was studied by Rheinboldt in~\cite[Theorem 4.3]{rheinboldt_unified_convergence_theory_1968}.
Below, we state his results, adapting them to functions in Euclidean spaces with one unique root.

\begin{theorem}
\label{thm: rheinboldt}
		Let $F:  \reals^{k} \mapsto \reals^{k} $ be a differentiable function, $F'(x)$ be the $k \times k$ Jacobian matrix of $F(x)$, 
		 $H:  \reals^{k} \mapsto \reals^{k}$ be a linear operator,
		 and $\xstep{0} \in \reals^{k}$ be an initial approximation of the iterative process~(\ref{eq: linear op step function}) for $F(x) = 0$. 
		Assume that: 
		\begin{description}
			\item{\emph{(1)}} $H(\xstep{0})$ is invertible and $\norm{H(\xstep{0})^{-1}} \le \beta$ 
			\item{\emph{(2)}}   $\norm{H(\xstep{0})^{-1} \, F(\xstep{0})} \le \eta$ 
			\item{\emph{(3)}} $\norm{F'(x)-F'(y)} \le L \norm{x-y}$ for $x,y \in  \reals^{k}$
			\item{\emph{(4)}} $\norm{H(x)-H(\xstep{0})} \le K \norm{x - \xstep{0}}$ for $x \in  \reals^{k}$
			\item{\emph{(5)}} $\norm{F'(x)-H(x)} \le \delta_0 + \delta_1 \norm{x - \xstep{0}}$ for $x \in \reals^{k}$ 
		\end{description}
		for some $\beta,\eta,L,K >0$ and $\delta_0, \delta_1 \ge 0$.
		
		With these assumptions, if $\beta \delta_0 < 1$ and $h =   \beta \eta L \, \max (1, (K + \delta_1)/L)/(1 - \beta \delta_0)^2 \le \frac{1}{2}$, 
		then the sequence $\braced{\xstep{t}}$ generated by equation~(\ref{eq: linear op step function}) starting at $\xstep{0}$ is well-defined, 
		contained in the ball $B(\xstep{0}, r)$ for $r = \eta (1-\sqrt{1-2h})/h(1- \beta \delta_0)$, 
		and converges linearly to the unique solution $\xop$ of $F(x) = 0$.
\end{theorem}

\smallskip
We apply Theorem~\ref{thm: rheinboldt} to Algorithm~HC-loop for a given flow graph $G$ in a similar manner to our use of Theorem~\ref{thm: kantorovich} for Algorithm~NR-loop. 
The bound in~(3) involving $L$ remains consistent with those outlined in Section~\ref{sec: analysis of Newton method}. 
For (1), the computation of the inverse of the diagonal matrix $H(\xstep{0})$ is straightforward: 
\begin{equation*}
	\beta = \norm{H(\xstep{0})^{-1}} \;=\; \max_{1 \le c \le k} \bracked{\frac{\partial f_c}{\partial x_c} (\xstep{0})}^{-1}
\end{equation*} 

Similarly, the bound in (2) can be taken as $\eta = \norm{H(\xstep{0})^{-1} \, F(\xstep{0})}$. 
For (4) and (5), the estimates for $K$, $\delta_0$, and $\delta_1$ depend on the cycle basis $C$, similar to $L$. 
We show the general estimates of these constants in the following claims. 

\begin{claim}
\label{claim: general bound on K}
	$\norm{H(x)-H(\xstep{0})} \le 2 (\ell - k + 1) \, \norm{x- \xstep{0}}$
\end{claim}

\smallskip
The derivation for this bound is as follows:
\begin{align*}
    	\norm{H(x)-H(\xstep{0})}  \;&= \; \max_{1 \le c \le k} \absvalueBig{\frac{\partial f_c}{\partial x_c} (x) -  \frac{\partial f_c}{\partial x_c} (\xstep{0}) } 
		\\
		\;&=\;  \max_{1 \le c \le k}  2  \absvalueBig{ \sum_{e=1}^{m} A_{ec}^2  \, \parendBig{\absvalue{\psi_e +  \sum_{i=1}^{k} A_{ei} \, x_i} - \absvalue{\psi_e+ \sum_{i=1}^{k} A_{ei} \, \xstep{0}_i} }}
		\\
		\;&\le\; \max_{1 \le c \le k}  2  \sum_{e=1}^{m}  A_{ec}^2 \,  \sum_{i=1}^{k}  \absvalue{A_{ei}}  \, \norm{x - \xstep{0} }
		\\
    		\;&=\; 2 (\ell - k +1) \, \norm{x- \xstep{0}} 
\end{align*}
because $\sum_{e=1}^{m} \sum_{i=1}^{k}  A_{ec}^2 \absvalue{A_{ei}}  \le \ell - k + 1$.

\begin{claim}
\label{claim: general bound on delta}
	$\norm{F'(x)-H(x)} \le   \delta_0 +  \delta_ 1 \, \norm{x- \xstep{0}}$  
	where $\delta_0 =  2(\ell - k -2) \,(\maxinitialflow +  k \norm {\xstep{0}} ) $, $ \delta_1 =2 k \, (\ell - k - 2)$, 
	and $\maxinitialflow = \max \limits_{1 \le e \le m} \absvalue{\psi_e}$. 
\end{claim}

\smallskip
The derivation for this bound is as follows:
\begin{align*}
    	\norm{F'(x)-H(x)}  \;&= \; \max_{1 \le c \le k} \sum_{d \ne c} \absvalueBig{\frac{\partial f_c}{\partial x_d} (x) } 
				\\
				\;&= \; \max_{1 \le c \le k} \sum_{d \neq c} 2 \, \absvalueBig{  \sum_{e=1}^{m} A_{ec} \, A_{ed} \, \absvalue{\psi_e + \sum_{i=1}^{k} A_{ei} \, x_i} } 
				\\
				\;& \le \;\max_{1 \le c \le k} 2 \, \sum_{d \neq c} \sum_{e=1}^{m}  \absvalue{A_{ec}} \, \absvalue{A_{ed}} \parendbig{ \maxinitialflow + k \norm{x}} 
				\\
				\;& \le \; 2(\ell - k -2) \, \parendbig{ \maxinitialflow + k \norm{\xstep{0}}  + k \norm{x - \xstep{0}}}
				\\
				\;& \le \;2(\ell - k -2)  \, (\maxinitialflow  + k \norm{\xstep{0}} \,)  + 2k\,(\ell - k -2) \, \norm{x - \xstep{0}} 
\end{align*}
because $\sum_{d \ne c} \sum_{e=1}^{m} \absvalue{A_{ec}} \, \absvalue{A_{ed}} \le \ell - k + 1  - 3 = \ell - k - 2$.
(This uses the fact that each cycle $c$ has at least 3 edges.)


The following theorem states the sufficient condition for linear convergence of Algorithm HC-loop 
on loop equation $F: \reals^k \mapsto \reals^k$, for a flow graph $G = (V,E)$ with a given cycle basis $C$ of total length $\ell$, 
a reference flow $\psi$ with the maximum absolute flow value flow $\maxinitialflow$, and an initial solution $\xstep{0}$ for which $H(x^{(0)})$ is invertible:

\begin{theorem}[Convergence conditions of Algorithm~HC-loop] 
\label{thm: hc convergence}
	Let  $\beta = \norm{H(\xstep{0})^{-1}}$, $\eta = \norm{H(\xstep{0})^{-1} \, F(\xstep{0}}$, $K = 2(\ell - k + 1)$, 
	$L = 2k \, (\ell - k + 1)$, $\delta_0 = 2 (\ell - k - 2) \, \parendbig{\maxinitialflow  + k \norm{\xstep{0}} }$, and $\delta_1 = 2k \, (\ell - k -2)$. 
	If $\beta \delta_0 < 1$ and $\beta \eta L \max(1, (K + \delta_1)/L) / (1-\beta \delta_0)^2 \le \frac{1}{2}$, then Algorithm HC-loop converges to the unique solution 
	$\xop$ of $F(x) = 0$ with linear order. 
\end{theorem}

We remark, similar to Section~\ref{sec: analysis of Newton method}, 
more specific estimates of the constants $K, \delta_0$ and $\delta_1$ can be obtained by substituting the value for $\ell$ 
corresponding to the cycle basis used in the algorithm. 

In the case where $G$ is planar and the face cycle basis is used, the constant $K$ can be improved to $4n$. 
This bound can be derived as follows: 
 \begin{align*}
    	\norm{H(x)-H(\xstep{0})} \;&\le\; \max_{1 \le c \le k}  2 \,   \sum_{e=1}^{m}  A_{ec}^2 \,  \sum_{i=1}^{k}  \absvalue{A_{ei}}  \, \norm{x - \xstep{0} }
		\\
    		\;&\le\; \max_{1 \le c \le k}  2 \,  \sum_{e=1}^{m} A_{ec}^2  \cdot 2 \,    \norm{x - \xstep{0}}
		\\
    		\;&\le \; 4 n \, \norm{x- \xstep{0}} 
\end{align*}

Furthermore, utilizing the property that for any edge $e$, $\sum_{i=1}^{k} \absvalue{ A_{ei} } \le 2$, 
the constants $\delta_0$ and $\delta_1$ can also be improved to $2(\ell - k - 2) \parendbig{\maxinitialflow + 2 \norm{\xstep{0}}}$ and $4(\ell - k - 2)$,
respectively. 

\smallskip
\myparagraph{Lower bound on convergence}
We now demonstrate that, in general, the guaranteed local convergence rate of Algorithm~HC-loop
is not better than linear. To illustrate this, consider the flow graph $G$ with 2 vertices, 3 edges, 1 inlet and 1 outlet. 
The cycle basis consists of two cycles, whose their clockwise directions are shown in the figure below. 
\begin{figure}[ht]
	\begin{center}
		\includegraphics[width = 3.5in]{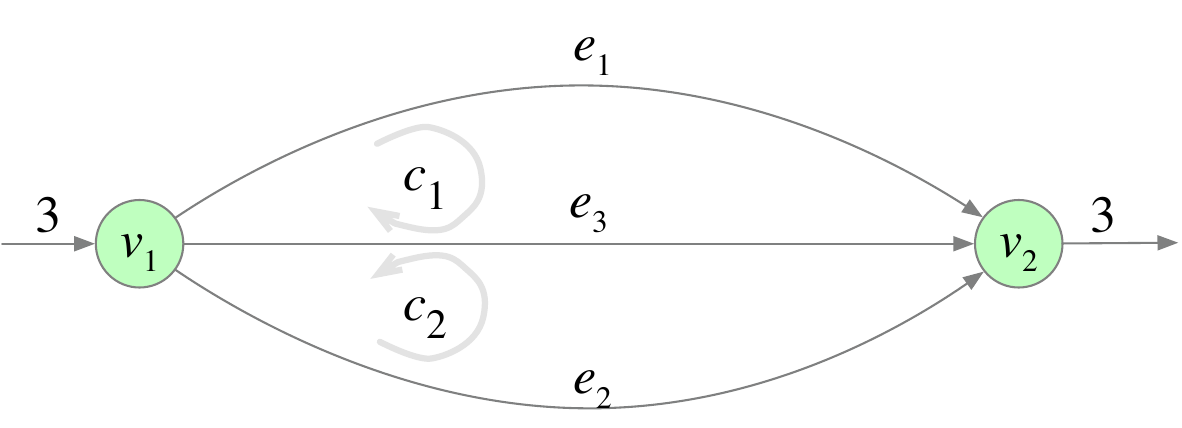}
		\label{fig: graph 2 cycles}
	\end{center}
\end{figure}

Let the reference flow be $\psi = [0,0,3]$. The loop equations are:
\begin{align*}
	f_1(x) \;&=\; \signsquare{x_1} - \signsquare{3 - x_1 - x_2}
	\\
	f_2(x) \;&=\; \signsquare{x_2} - \signsquare{3 - x_1 - x_2}
\end{align*}
The solution is $\xop_1 = \xop_2 = 1$.  
We are interested in the case where the initial adjustment $\xstep{0}$ is sufficiently close to the solution, 
particularly when $\xstep{0}_1 = \xstep{0}_2 =  1 \, \pm \, \epsilon$ for some $0 < \epsilon \ll 1$. 
The first adjustment is:
\begin{equation*}
	\xstep{1}_1 = \xstep{1}_2 = 1 \pm  \epsilon \mp \frac{3 \epsilon \, (2 \, \pm \, \epsilon)}{2(2 \, \pm \, \epsilon)} = 1 \, \pm \, \epsilon/2
\end{equation*}
Thus, $\norm{\,\xstep{1} - \xop \,} =  \epsilon/2 = \half \, \norm{\,\xstep{0} - \xop \,}$. The convergence rate is exactly $\half$.

\section{Conclusion}
\label{sec: conclusion}


We expressed the pipe network analysis problem as a root-finding problem for a system of non-linear equations 
and we discussed the local convergence conditions of Algorithms NR-loop and HC-loop (that is the loop variants of
the Newton-Raphson and HardyCross algorithms). 
We showed that Algorithm NR-loop achieves quadratic local convergence, while HC-loop ensures only linear local convergence. 
These findings align with experimental results in the literature. 
Additionally, we demonstrated that by using the face cycle bases, the convergence conditions for both algorithms are 
strengthened for the important case of planar graphs.


\bibliographystyle{plain}

\bibliography{convergence_analysis_of_pipe_network_references}


\appendix

\section{Errors in~\cite{altman_convergence_Newton_nonlinear_1994}}
\label{sec: altman errors}


Altman and Boulos~\cite{altman_convergence_Newton_nonlinear_1994} attempted to provide local convergence conditions of
Algorithm~NR-loop and Algorithm~NR-flow for solving flow equations. 
However, their analysis has two errors that make their bounds on convergence invalid.
The first one is an algebraic mistake when substituting the variables in~\cite[ eq.~(13)]{altman_convergence_Newton_nonlinear_1994}: 
\begin{align*}
	x_1  \;&=\; \eta_{11}  x_{t+1} + \eta_{12}  x_{t+2} + \cdots + \eta_{1t}  x_{e} + a_1
	\\
	&\vdots
	\\
	x_t \;&=\; \eta_{t1}  x_{t+1} + \eta_{t2}  x_{t+2} + \cdots + \eta_{tt}  x_{e} + a_t
\end{align*}
into~\cite[ eq.~(12)]{altman_convergence_Newton_nonlinear_1994}: 
\begin{align*}
	\gamma_{11} \xi_1 x_1^2 + \gamma_{12} \xi_2 x_2^2 + \cdots + \gamma_{1e} \xi_e x_e^2 + \Phi_1 \;&=\; 0 
	\\
	&\vdots
	\\
	\gamma_{m1} \xi_1 x_1^2 + \gamma_{m2} \xi_2 x_2^2 + \cdots + \gamma_{me} \xi_e x_e^2 + \Phi_m \;&=\; 0 
\end{align*}
(The indices in the last equation in~\cite{altman_convergence_Newton_nonlinear_1994} were also incorrectly
given as $n$ instead of $m$. We corrected these above.)
With a change of variables from $x_{t+1}, x_{t+2}, \cdots, x_e$ to $x_1, x_2, \cdots, x_l$, they obtained ~\cite[ eq.~(14)]{altman_convergence_Newton_nonlinear_1994}:
\begin{align*}
	f_1 \;&=\; p_{11} x_1^2 + p_{12} x_2^2 + \cdots + p_{1l} x_l^2 + b_{11} x_1 + b_{12} x_2 + \cdots + b_{1l} x_l + d_1 = 0
	\\
	&\vdots
	\\
	f_l \;&=\; p_{l1} x_1^2 + p_{l2} x_2^2 + \cdots + p_{1l} x_l^2 + b_{l1} x_1 + b_{l2} x_2 + \cdots + b_{ll} x_l + d_l = 0
\end{align*}
All the terms $x_i x_j$ where $i \ne j$ do not appear in the above equations 
which affect the computation of the partial derivatives $\partial f_i/\partial x_j$ later on in their analysis. 

The other mistake involves the sign issue in the energy equation for each cycle. 
The authors used a fixed indicator $\gamma_{wu}$ for the orientation of flow in cycle $w$ in~\cite[eq.~(6)]{altman_convergence_Newton_nonlinear_1994}:
\begin{align*}
	\Phi_w + \sum_{u=1}^{e} \gamma_{wu} \xi_u Q_u^2 = 0
\end{align*}
This would be true if all flow directions remained the same, in other words, if the sign of each $Q_u$ was constant through the whole
iterative process.  However, it is not guaranteed. 
Thus, when $Q_u$ flips sign, $\gamma_{wu}$ need to be changed accordingly for the energy equation to be valid.


\section{An Example for Non-Convergece of Algorithm NR-node}
\label{sec: oscillation example}
\begin{figure}[h]
	\begin{center}
		\includegraphics[width = 3.5in]{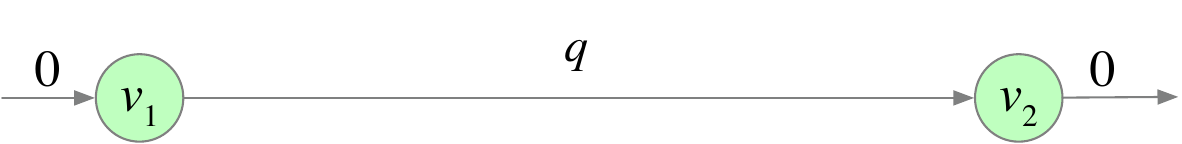}
		\label{fig: oscillation}
	\end{center}
\end{figure}

Shamir mentioned the issue with non-convergence encountered in certain systems  in~\cite{shamir_water_distribution_analysis_1968}. 
We provide a simple flow graph in which Algorithm~NR-node enters an infinite loop. 
Using similar setup for the node method as in~\cite{shamir_water_distribution_analysis_1968}, 
the unknowns are pressure $p_v$ at node $v$ for  $v = 1,2,\cdots, n$, and the system of equations includes
flow conservation equations for each node of the form:
\begin{equation}
	\label{eqn: nr-node function}
	f_v(p_1,p_2,\cdots,p_n) = w_v + \sum \limits_{e = (u,v) \in E} D_{ev} \, q_e
	= w_v +\sum \limits_{e = (u,v) \in E} D_{ev} \frac{p_u - p_v}{\sqrt{|p_u - p_v|}} \;=\; 0
\end{equation} 

The below example demonstrates an oscillation case of Algorithm~HR-node for solving the described node method equation. 
Consider the simple flow graph $G$ with only two vertices $v_1$ and $v_2$ connected by a single edge 
as shown in the figure below. Set the external flow on the two vertices to be
$0$ and the corresponding pressures to be  $p_1 = x\neq 0$ and $p_2 = 0$. 

The equation~\eqref{eqn: nr-node function} is then  $f(x) = \frac{- x}{\sqrt{|x|}} = 0$.
The iteration step is:
$x - \frac{f(x)}{f'(x)} = x - \frac{x \cdot 2 \sqrt{|x|}}{\sqrt{|x|}} = -x$.
So for any initial value $x \neq 0$, the algorithm will oscillates between $x$ and $-x$.

\end{document}